\documentclass[letterpaper]{article}
\usepackage[preprint]{preprint}

\usepackage[T1]{fontenc}
\usepackage[hyphens]{url}
\usepackage{graphicx}
\usepackage{natbib}
\usepackage{caption}
\usepackage{amsmath}
\usepackage{multirow}
\usepackage{booktabs}

\title{ParticleGen: A Multi-Agent System for Particle Effects Generation}
\author{
  Junhao Zhuge\textsuperscript{1},
  Junyi Yang\textsuperscript{1},
  Yuqing Wang\textsuperscript{1},
  Kangzhan Wang\textsuperscript{1},
  Sipeng Yang\textsuperscript{2},
  Xiaogang Jin\textsuperscript{1}\corresponding
}
\affiliations{
  \textsuperscript{1}State Key Lab of CAD\&CG, Zhejiang University, China\\
  \textsuperscript{2}Hangzhou Research Institute of AI and Holographic Technology, China\\
  22421251@zju.edu.cn, 22560234@zju.edu.cn, 2430776967@qq.com,\\
  3230105476@zju.edu.cn, sipengyang@zjuqx.com, jin@cad.zju.edu.cn
}
\date{}

\begin{document}

\maketitle

\begin{abstract}
Particle systems are widely used in digital entertainment to create dynamic scene elements and visual effects. However, authoring high-quality particle effects remains labor-intensive and demands specialized expertise, requiring practitioners to navigate complex procedural rules and high-dimensional parameter spaces. Recent large language models (LLMs) enable users to specify particle effects through natural language, yet reliably translating high-level creative intent into executable procedural logic and low-level parameters remains difficult.
In this work, we present a multi-agent framework for the from-scratch synthesis of structured and editable particle systems from natural language descriptions. Given a text prompt, our framework first generates an initial particle configuration through a decoupled planning and parameterization pipeline, and then iteratively improves the result based on rendered feedback. To support precise and targeted adjustments, we further introduce a diagnostic mechanism that links observed visual artifacts to their underlying procedural causes. We validate our approach in Unreal Engine 5's Niagara system across a diverse set of scenarios, including elemental spells, dynamic natural phenomena, and fireworks. Quantitative and qualitative evaluations show that our method achieves high semantic fidelity and visual quality. By directly synthesizing structured particle simulation logic, our framework reduces the technical barrier to particle effect authoring and improves the efficiency of creative iteration.
\end{abstract}

\begin{figure*}[t]
  \includegraphics[width=\textwidth]{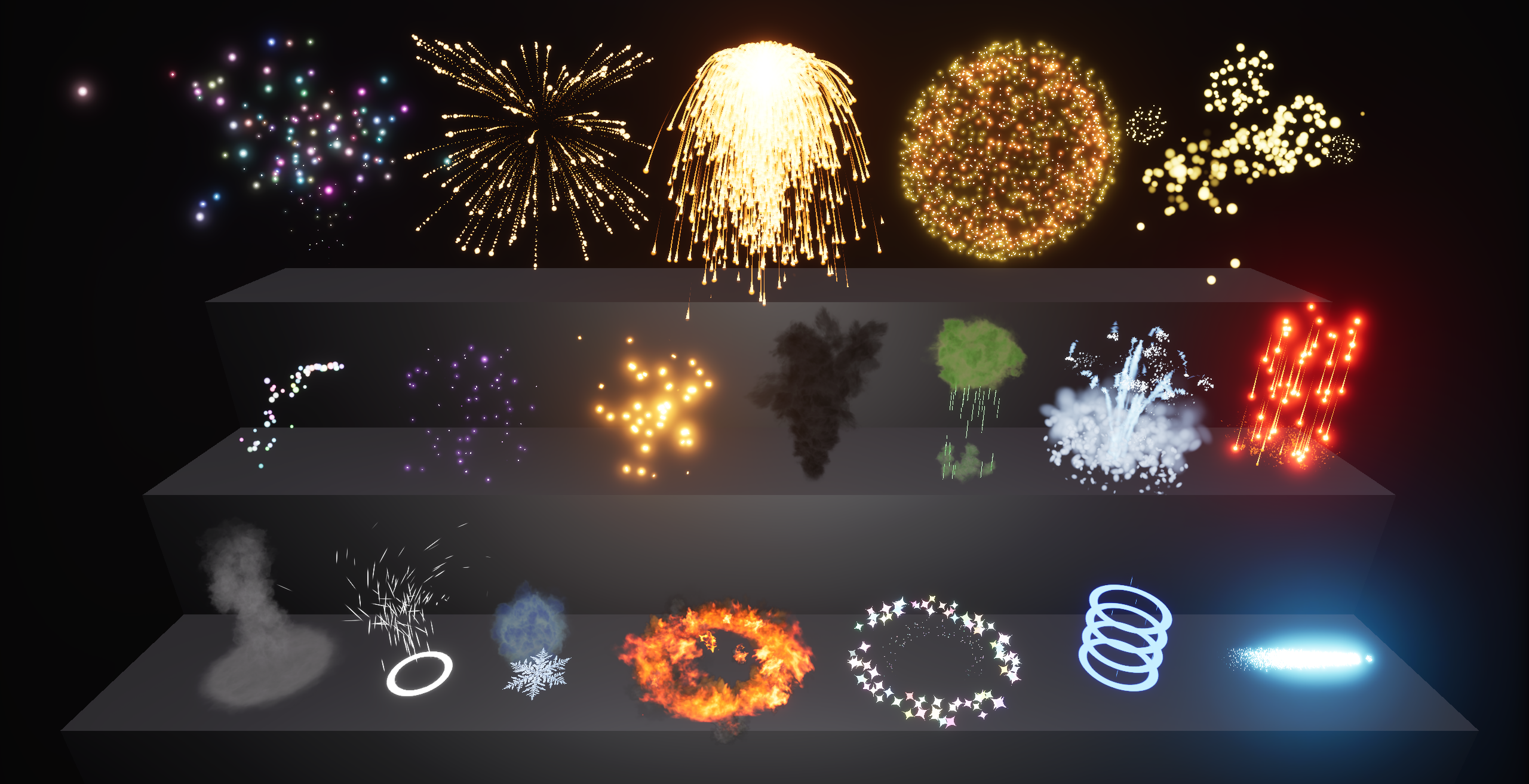}
  \caption{\textbf{A gallery of diverse particle effects synthesized from natural language descriptions by our framework.} The proposed method enables the automated creation of high-fidelity systems encompassing a vast array of visual styles and physical behaviors. The showcased examples include natural phenomena, fireworks, and magical motifs, highlighting the versatility and extensibility of our pipeline.}
  \label{fig:teaser}
\end{figure*}

\section{Introduction}

Particle effects are widely used in video games, animated films, and short-form video content to simulate complex dynamic phenomena, such as rain, smoke, explosions, and magical effects. However, crafting high-quality particle effects remains difficult for non-expert users, as it requires technical knowledge of particle system design and visual aesthetic sensitivity. Even for experienced technical artists, navigating the high-dimensional parameter spaces of these systems necessitates a labor-intensive and cumbersome optimization process. Consequently, there is a pressing need for automated solutions that can simplify and accelerate particle effect creation.

Despite this demand, automatic generation of high-quality particle effects remains underexplored. A recent work, KinemaFX~\cite{zhang2025kinemafx}, synthesizes particle effects from natural-language descriptions, but its \textit{retrieval-based} formulation constrains the output space to a predefined effect database. As a result, it has limited ability to create novel effects or represent complex procedural dependencies beyond simple temporal sequencing. Furthermore, the method still requires substantial user interaction, limiting authoring efficiency in practical workflows.

Recent large language models (LLMs) have shown strong capabilities in code generation, structured output, and multi-step reasoning. These capabilities suggest the possibility of synthesizing particle effects from scratch, rather than retrieving and recombining existing assets. However, directly generating particle systems with LLMs remains highly challenging. The authoring process involves high-dimensional parameters, complex procedural logic, and dependencies among simulation modules that must be coordinated to produce the intended visual effect. In addition, particle-system knowledge is often tied to specific implementations, such as engine-specific modules and parameter conventions, making supervised fine-tuning difficult to scale across different systems.

In this paper, we propose ParticleGen, a multi-agent system that autonomously synthesizes complex particle effects from natural-language descriptions. Our framework begins with a two-stage initial generation phase: a \textit{Planner agent} first defines the compositional architecture and assigns functional responsibilities to individual emitters; a set of parallel \textit{Generator agents} then perform behavioral parameterization grounded in a modular technical knowledge base. The resulting structured text representation is subsequently translated into native engine assets and rendered into video sequences to facilitate visual evaluation. To reduce discrepancies between the generated effect and the user's intended description, we introduce a closed-loop iterative refinement stage. In this loop, a \textit{Critic agent} identifies spatio-temporal mismatches in the rendered video sequences to provide visual feedback to a \textit{Refiner agent}, which produces targeted modifications.
For this refinement process, we use retrieval-augmented generation (RAG)~\cite{lewis2020retrieval} as a diagnostic tool, which we term Diagnostic RAG (DRAG), to trace observed visual artifacts to their underlying procedural causes.

We demonstrate the efficacy of our framework through a systematic validation within the Unreal Engine 5 (UE5) \cite{ue5} Niagara system \cite{niagara}. To our knowledge, this is the first attempt to harness the generative capabilities of LLMs for the direct synthesis of 3D real-time particle effects with intricate simulation logic via structured representations, effectively distinguishing our work from retrieval-based paradigms.
Our contributions are summarized as follows:
\begin{itemize}
\item A multi-agent generative framework enabling the direct synthesis of native, fully editable real-time particle effects from text description.
\item A closed-loop iterative refinement mechanism specifically designed for complex particle systems, utilizing patch-based modifications and a selective rollback strategy augmented by DRAG.
\item A systematic validation within UE5's Niagara particle system, demonstrating the framework's capability to generate high-fidelity particle effects in a modern industrial pipeline.
\end{itemize}

\section{Related Work}
\subsection{Particle Effect Authoring}
Early attempts to simplify particle effect creation \cite{arora2019magicalhands, xie2020body2particles} established direct mappings between user inputs and a restricted set of simulation attributes. While providing intuitive control for targeted applications, their reliance on hard-coded mappings within a narrow behavioral scope positions them as specialized control interfaces rather than autonomous generative systems.

To move beyond straightforward parameter mapping, researchers have investigated search-based co-creation workflows. One approach leverages evolutionary algorithms to explore the parameter space via interactive visual selection, enabling precise tuning of fundamental parameters such as emission rate \cite{chueca2024search}. KinemaFX \cite{zhang2025kinemafx} advances this paradigm by shifting the focus from low-level parameters to pre-authored effect libraries, allowing users to iteratively assemble complex effects from retrieved candidates. While this retrieval-based approach offers high visual fidelity, its expressiveness remains inherently limited by the diversity and coverage of the underlying libraries.

Actionbrushes \cite{nai2026actionbrushes} represents a pioneering effort as the first work to utilize LLMs for the direct synthesis of particle system parameters from natural language prompts. However, the scope of this synthesis is restricted to 2D animation and a quite narrow set of simulation parameters, which limits its capacity to support the complex logical interdependencies between particles required for sophisticated visual effects, such as event-driven spawning or collision-based triggers. In contrast, our framework achieves the \textit{de novo} synthesis of engine-native particle systems. By leveraging a multi-agent orchestration grounded in systematic technical documentation, we enable the generation of comprehensive behavioral parameters and intricate simulation logic.

\subsection{Agentic Workflow}
Early paradigms for autonomous generation focused on enhancing the performance of single LLM through structured strategies such as Chain-of-Thought \cite{wei2022chain}, ReAct \cite{yao2023react} and Plan-and-Solve \cite{wang2023plan}, which encourage explicit reasoning steps before action execution. While effective for moderate complexity \cite{li2024anim, you2025designmanager, hwang2025does}, these single-agent approaches often suffer from hallucination \cite{ji2023survey, huang2025survey} and reasoning overload in highly intricate generative tasks. Building on these foundations, recent research has evolved toward agentic workflows \cite{hong2023metagpt, xu2024magic, wu2024autogen} that decompose intricate problems into specialized sub-tasks assigned to multiple coordinated agents. This multi-agent approach has demonstrated significant potential across diverse domains, exemplified by applications in video editing \cite{sandoval2025editduet}, animation making \cite{shi2025animaker}, nonverbal behavior generation \cite{zhang2025social}, poster authoring \cite{pangpaper2poster}, game play \cite{wei2026facul}, and game creation \cite{jiang2026opengame}. In these systems, the reasoning burden is distributed among specialized agents to ensure higher consistency and technical accuracy. Furthermore, due to the inherent difficulty of these tasks, these workflows often incorporate iterative refinement mechanisms \cite{shinn2023reflexion, madaan2023self} that utilize synthetic feedback or scoring signals to progressively enhance generative quality. In parallel, latent-space collaboration methods \cite{zou2025latent, du2025enabling} explore bypassing discrete tokenization to eliminate information loss during agent communication. However, they require direct access to model internals only feasible with open-source models, whose capabilities are currently insufficient for the complex reasoning in our particle synthesis task.

\section{Method}
\subsection{Framework Overview}
\begin{figure*}

\includegraphics[width=\textwidth]{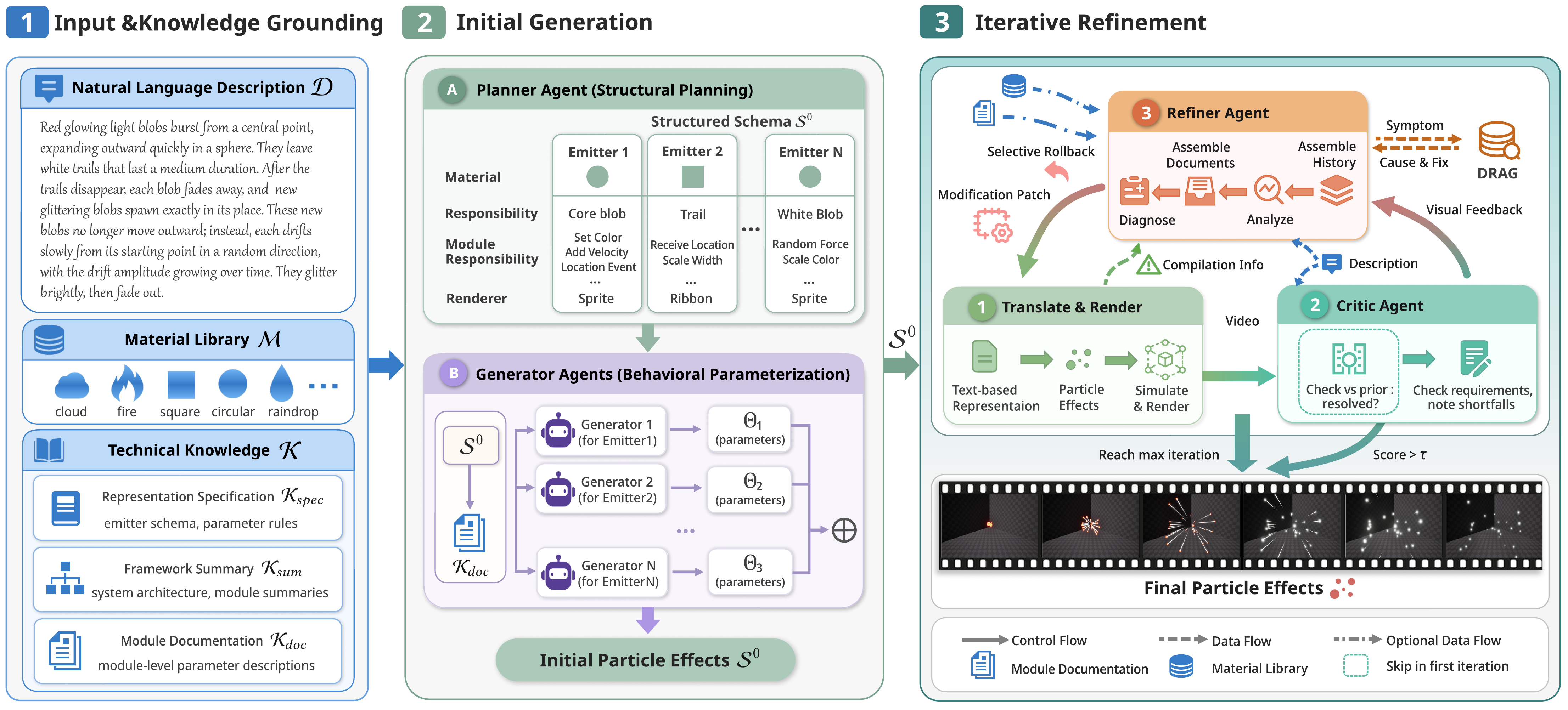}
  \vspace{-15pt}
  \caption{\textbf{Overview of our multi-agent framework for text-driven particle effect generation.}
  (1) \textbf{Input \& Knowledge Grounding} serves as the foundation, providing the user's natural language description $\mathcal{D}$ of desired particle effects, a pre-authored material library $\mathcal{M}$ with text description, and domain-specific technical knowledge.
  Driven by these grounded inputs, the generative pipeline executes in two primary phases:
  (2) \textbf{Initial Generation}: A \textit{Planner agent} first creates a structured schema $\mathcal{S}^0$, followed by multiple \textit{Generator agents} that perform behavioral parameterization in parallel.
  (3) \textbf{Iterative Refinement}: A closed-loop execution where a \textit{Critic agent} evaluates the rendered video, and a \textit{Refiner agent} performs automated diagnosis to iteratively patch the parameters until high-quality effects are achieved or the maximum iteration limit is reached.}
  \label{fig:pipe}
\end{figure*}
In this section, we describe our framework designed to bridge the gap between high-level artistic intent and the intricate parameterization of particle systems. Let $\mathcal{D}$ be a natural language description of the desired particle effect provided by the user, $\mathcal{M}$ be the detailed textual descriptions of a curated library of material assets covering their silhouette morphology, color characteristics, and temporal behaviors, and $\mathcal{K}$ represent the detailed technical documentation of the underlying particle framework. Each material $m \in \mathcal{M}$ is associated with a detailed textual description covering its morphology, chromaticity, and temporal behaviors. We formalize the creation of the effect as a generative synthesis task $\mathcal{G}$:
\begin{equation}
    \mathcal{G}: (\mathcal{D}, \mathcal{M}, \mathcal{K}) \rightarrow \mathcal{S}.
\end{equation}
The synthesized particle system $\mathcal{S}$ is composed of multiple particle emitters:
\begin{equation}
    \mathcal{S} = \{E_1, E_2, \dots, E_n\},
\end{equation}
where each emitter $E_i$ is defined as a triplet:
\begin{equation}
    E_i = \langle \Phi_i, \mathcal{R}_i, \mathbf{m}_i \rangle.
\end{equation}
Each emitter $E_i$ consists of the following components:
\begin{itemize}
    \item \textbf{Behavioral Modules ($\Phi_i$)}: A sequenced set of modules $\Phi_i = \{\phi_1, \dots, \phi_k\}$ responsible for particle behavior. Each module $\phi$ is defined by a configuration parameter set $\theta$. These parameters collectively dictate the evolution of the particle state across successive simulation frames.
    \item \textbf{Renderer Configuration ($\mathcal{R}_i$)}: The visual output logic, including the renderer type (e.g., Sprite, Ribbon), orientation modes, etc.
    \item \textbf{Material ($\mathbf{m}_i \in \mathcal{M}$)}: The material picked from the available library $\mathcal{M}$ that best matches the semantic context of $E_i$.
\end{itemize}

The technical documentation $\mathcal{K}$ provides the necessary grounding for the behavioral modules and the overall system structure. We formalize $\mathcal{K}$ as a structured knowledge base comprising three distinct components:
\begin{enumerate}
    \item \textbf{Representation Specification ($\mathcal{K}_{spec}$)}: The syntax and structural constraints governing the text-based representation of the particle system.
    \item \textbf{Framework Summary ($\mathcal{K}_{sum}$)}: A high-level overview of the particle system framework, encompassing functional summaries of available modules, renderer types, and their global capabilities. This component is utilized to guide macro-architectural decisions during the planning stage.
    \item \textbf{Behavioral Module Documentation ($\mathcal{K}_{doc}$)}: An indexed knowledge base of detailed parameter specifications. We define $\mathcal{K}_{doc}$ as a queryable document such that for a given module $\sigma$, $\mathcal{K}_{doc}(\sigma)$ retrieves the technical parameters and physical meanings.
\end{enumerate}

To solve the synthesis task $\mathcal{G}$, we propose a multi-agent framework that first constructs an initial particle system $\mathcal{S}$ through a two stage generative process. Subsequently, the system enters an iterative refinement stage where a \textit{Refiner agent} utilizes visual guidance from a \textit{Critic agent} to optimize the parameters via modification patches. This guidance is produced by the \textit{Critic agent} through the evaluation of video sequences, which are rendered by translating the text-based representation of the particle system into the particle engine's native binary format.
\subsection{Initial Generation}
The initial generation of the particle system $\mathcal{S}$ is conducted through a two-stage process that transitions from high-level structural orchestration to low-level parameter synthesis.

\paragraph{Structural Planning.}
In the first stage, a \textit{Planner agent} acts as a system architect to determine the overall schema of $\mathcal{S}$ based on the user description $\mathcal{D}$, the material library $\mathcal{M}$, and the technical documentation $\mathcal{K}$. This structural orchestration is formalized as a generation process $\mathcal{G}_{plan}$ performed by the \textit{Planner agent}:
\begin{equation}
    \mathcal{S}^0 = \mathcal{G}_{plan}(\mathcal{D}, \mathcal{M}, \mathcal{K}_{spec}, \mathcal{K}_{sum}).
\end{equation}
The synthesized schema $\mathcal{S}^0$ consists of $n$ discrete skeletal emitters:
\begin{equation}
    \mathcal{S}^0 = \{E_1^0, E_2^0, \dots, E_n^0\},
\end{equation}
where each skeletal emitter $E_i^0$ is defined as the quadruple:
\begin{equation}
    E_i^0 = \langle \mathbf{m}_i, \mathcal{R}_i, \hat{\Phi}_i, \Omega_i \rangle.
\end{equation}
In this formulation, $\Omega_i$ represents the emitter-level functional responsibility, while $\hat{\Phi}_i$ denotes the skeletal module sequence that defines the emitter behavioral structure. $\hat{\Phi}_i$ is further formalized as an ordered sequence of module-responsibility pairs:
\begin{equation}
    \hat{\Phi}_i = ( \langle \sigma_{i,1}, \omega_{i,1} \rangle, \langle \sigma_{i,2}, \omega_{i,2} \rangle, \dots, \langle \sigma_{i,k}, \omega_{i,k} \rangle ),
\end{equation}
where $\sigma_{i,j}$ represents the uninitialized module identifier retrieved from $\mathcal{K}_{sum}$ and $\omega_{i,j}$ denotes the specific functional responsibility assigned to that module. These functional responsibilities provide the semantic context required for inter-emitter dependency awareness and subsequent parameterization. Material $\mathbf{m}_i$ is selected by prioritizing silhouette morphology and intrinsic temporal behavior during this process. This prevents module-mutable attributes like color, opacity or scale from biasing the selection. At the conclusion of this stage, the structural layout of each emitter is established, while the configuration parameter sets $\theta$ remain uninitialized.

\paragraph{Behavioral Parameterization.}
Once the structural blueprint $\mathcal{S}^0$ is finalized, the process proceeds to behavioral parameterization. To navigate the high-dimensional parameter space and avoid attention dilution, the framework instantiates multiple \textit{Generator agents} in parallel, each dedicated to a single emitter $E_i^0$. This parallelization serves a dual purpose: it improves computational efficiency while also managing the context window. For each emitter, the assigned agent synthesizes the uninitialized parameter sets for the entire sequence $\hat{\Phi}_i$:
\begin{equation}
    \Theta_i = \mathcal{G}_{gen}(\mathcal{D}, E_i^0, \mathcal{K}_{doc}|_{\hat{\Phi}_i}),
\end{equation}
where $\Theta_i = (\theta_{i,1}, \theta_{i,2}, \dots, \theta_{i,k})$ represents the collection of synthesized parameters for all modules in the sequence. The term $\mathcal{K}_{doc}|_{\hat{\Phi}_i}$ denotes the subset of technical documentation restricted to the specific module types $\sigma \in \hat{\Phi}_i$, thereby reducing contextual noise. By interpreting the nuanced dynamics in $\mathcal{D}$ alongside the functional responsibilities $\Omega_i$ and $\omega_{i,j}$ within $E_i^0$, the agent maps linguistic cues to precise numerical values, procedural curves, or cross-module attribute bindings. The resulting instantiated emitter is formalized as $E_i = \langle \Phi_i, \mathcal{R}_i, \mathbf{m}_i \rangle$, where the fully parameterized module sequence is $\Phi_i = ( \phi_{i,1}(\theta_{i,1}), \dots, \phi_{i,k}(\theta_{i,k}) )$. This stage concludes with the formation of the complete particle system $\mathcal{S} = \{E_1, \dots, E_n\}$.

\subsection{Iterative Refinement}
Following the construction of the initial system $\mathcal{S}_0$, the framework enters a feedback-driven closed-loop optimization stage. This process is formalized as an iterative sequence of system states $\{\mathcal{S}_t\}_{t=1}^T$, where each iteration aims to minimize the visual discrepancy between the current simulation and the user intent $\mathcal{D}$.

\subsubsection{Visual Guidance via Criticism.}
Each refinement cycle begins with the translation of the text-based representation $\mathcal{S}_t$ into a rendered video sequence $V_t$:
\begin{equation}
    V_t = \text{Render}(\text{Translate}(\mathcal{S}_t)).
\end{equation}
Specifically, \textit{Translate} converts the schema $\mathcal{S}_t$ into executable particle system binaries, while \textit{Render} produces the video sequence $V_t$  by simulating the particles within a predefined scene.
The \textit{Critic agent} then performs a visual evaluation of $V_t$ against the artistic intent $\mathcal{D}$. For iterations $t > 0$, this process is inherently comparative, assessing whether identified artifacts have been mitigated or exacerbated relative to $V_{t-1}$. We formalize the generation of the visual alignment score $s_t$ and the visual guidance $G_t$ as:
\begin{equation}
    (G_t, s_t) =
    \begin{cases}
        \mathcal{G}_{critic}(V_0, \mathcal{D}, \emptyset), & t = 0, \\
        \mathcal{G}_{critic}(V_t, \mathcal{D}, V_{t-1}), & t > 0.
    \end{cases}
\end{equation}
The symbol $\emptyset$ denotes the absence of a prior video sequence during the initial iteration. In this formulation, $s_t \in [0, 1]$ represents the semantic alignment between the visual output and $\mathcal{D}$. The term $G_t$ denotes the visual guidance, which explicitly identifies mismatches in spatio-temporal dynamics and visual appearance. These observations serve as a qualitative bridge, translating observed visual flaws into actionable suggestions for the \textit{Refiner agent}.

\subsubsection{Parameter Optimization via Modification Patches.}
The \textit{Refiner agent} performs targeted updates by synthesizing a structured modification patch $\Delta \mathcal{J}_t$ based on a reconstructed context $\mathcal{C}_t$. This reconstruction is necessitated by the verbose nature of the particle system's structured representation. The cumulative inclusion of historical system states would lead to a rapid and linear expansion of the context window, introducing significant contextual noise that degrades the agent's reasoning performance.

The context is reconstructed as $\mathcal{C}_t = \{ G_{0:t}, \Delta \mathcal{J}_{0:t-1}, \mathcal{S}_t, \mathcal{E}_t \}$, which comprises the current and historical visual guidance, the sequence of previous patches, the current system state, and the compilation feedback $\mathcal{E}_t$, which is from the translator to rectify technical violations that may have arisen during the initial generation of the system or the subsequent parameterization of parameter sets $\theta$. The inclusion of historical trace allows the agent to evaluate the efficacy of prior adjustments, enabling it to determine whether to persist with the current technical trajectory or pivot its strategy based on observed outcomes.

Utilizing a \textit{Reasoning and Acting} (ReAct) \cite{yao2023react} paradigm, the \textit{Refiner agent} invokes targeted queries to both $\mathcal{K}_{doc}$ and the material library $\mathcal{M}$ whenever $\mathcal{C}_t$ provides insufficient technical grounding for the intended adjustments. The retrieved knowledge, denoted as $\mathcal{K}_{req} \subset (\mathcal{K}_{doc} \cup \mathcal{M})$, is combined with $\mathcal{C}_t$ and $\mathcal{K}_{sum}$ to synthesize the modification patch:
\begin{equation}
    \Delta \mathcal{J}_t = \mathcal{G}_{refine}(\mathcal{C}_t, \mathcal{K}_{req}, \mathcal{K}_{sum}).
\end{equation}
The patch $\Delta \mathcal{J}_t$ encodes atomic operations to add or remove modules $\phi \in \Phi_i$, as well as to update materials $\mathbf{m}_i$, renderer configurations $\mathcal{R}_i$, and parameter sets $\theta$. This patch is applied to the current system state through a composition operator $\oplus$:
\begin{equation}
    \mathcal{S}_{t+1} = \mathcal{S}_t \oplus \Delta \mathcal{J}_t.
\end{equation}
By selectively modifying these emitter components, the agent achieves precise control over the structural organization and particle attributes without necessitating a full system reconstruction. To address the challenge of ambiguous causal attribution, where intricate dependencies often obscure the source of visual artifacts, we incorporate the DRAG mechanism. This component utilizes a knowledge base that pairs characteristic visual symptoms with their underlying root causes within the particle framework. By performing semantic matching between the visual guidance and this symptom-cause library, the \textit{Refiner} can accurately identify the parameters responsible for non-obvious simulation errors, keeping the refinement process grounded in established troubleshooting logic.

\begin{figure*}[t]
  \centering
  \includegraphics[width=\linewidth]{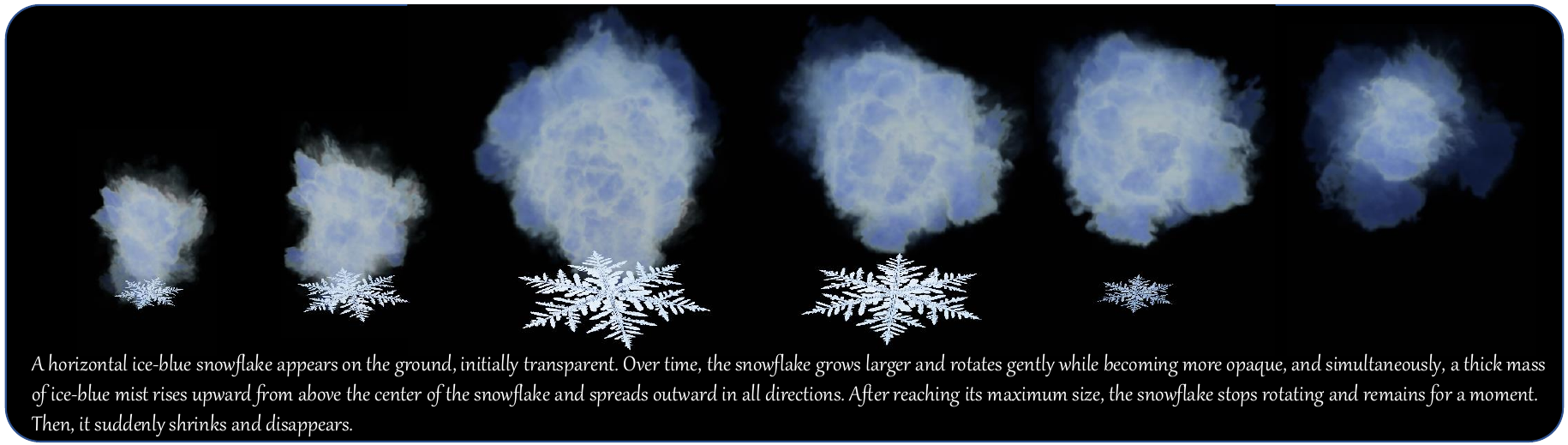}
  \includegraphics[width=\linewidth]{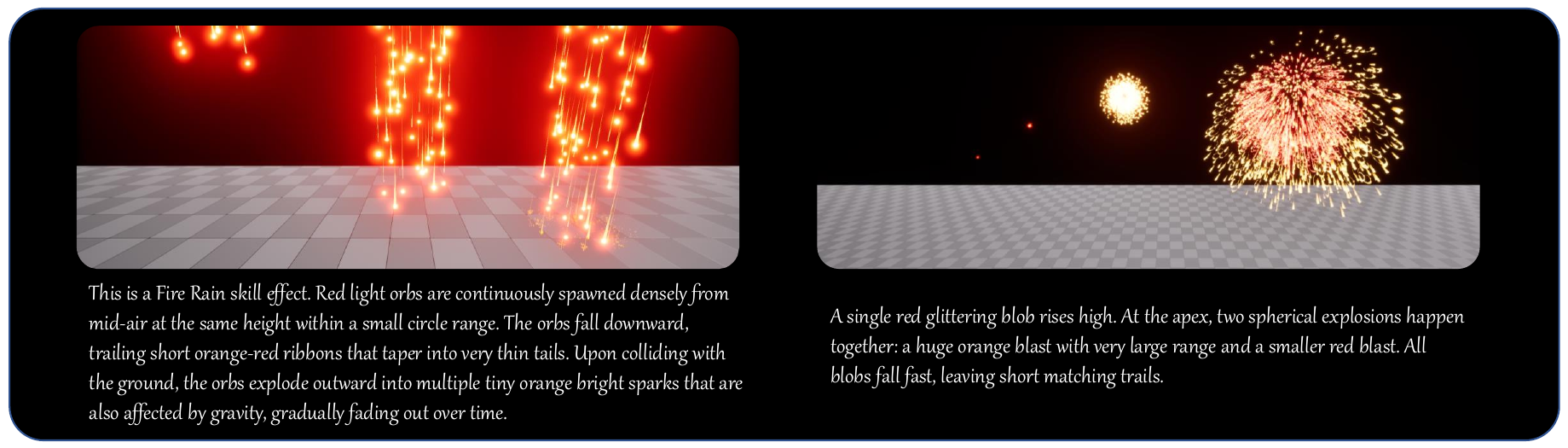}
  \vspace{-15pt}
  \caption{\textbf{Representative particle effects synthesized by our framework}. Each effect is displayed from left to right to illustrate the temporal evolution of the simulation. These results showcase the system's ability to generate diverse visual phenomena. More results are shown in Fig.~\ref{fig:gallery}.}
  \label{fig:showcase}
\end{figure*}

\subsubsection{Iterative Optimization Loop.}
The refinement process is conducted as a closed-loop optimization that terminates when the visual score $s_t$ given by the \textit{Critic} exceeds a predefined threshold $\tau$ or the iteration count reaches the maximum limit $T$. To ensure robust convergence, the framework implements a selective rollback mechanism facilitated by the objective-based structure of the modification patches. At each iteration, the \textit{Critic} provides comparative visual feedback by identifying which aspects of the simulation have improved or deteriorated relative to the previous state. Based on this feedback, the \textit{Refiner} assesses the efficacy of each objective-based segment within the prior modification patch $\Delta \mathcal{J}_{t-1}$. Any segments that failed to achieve their intended visual outcomes are signaled to rollback, allowing the system to revert unsuccessful changes while persisting with effective ones.

Recognizing that the refinement process may occasionally diverge or oscillate due to the stochastic nature of the underlying models, the framework preserves the system state $\mathcal{S}_t$ from each iteration. Rather than simply returning the final state $\mathcal{S}_T$, the framework performs a best-state selection, outputting the candidate that achieved the highest historical score to ensure optimal visual fidelity.

\section{Experiments}

\subsection{Implementation Details}
To demonstrate the feasibility and effectiveness of our framework, we conduct system validation using UE5 \cite{ue5} and its Niagara \cite{niagara} particle system as the target platform. We adopt JavaScript Object Notation (JSON) as the intermediate text-based representation for the synthesized particle systems, as LLMs demonstrate high proficiency in generating this format. To bridge the gap between structured representations and the engine's native binary format, we develop a new Unreal Engine plugin, \textit{FxConverter}, that translates JSON outputs into native engine assets.

We utilize GPT-5.4~\cite{gpt} as the underlying LLM in our framework, chosen for its long-context capability, strong reasoning performance, and reliable instruction following. These properties are important for handling the high-dimensional parameter spaces and structural constraints of particle systems.
The \textit{Critic agent} also uses GPT-5.4 for visual assessment; as the model lacks native video understanding capability, each rendered sequence is converted into a compact set of frames, with further details provided in the supplementary material.
We construct the technical knowledge base $\mathcal{K}$ to support the generation of engine-compatible particle-system specifications. Specifically, $\mathcal{K}_{spec}$ and $\mathcal{K}_{sum}$ are organized to match our text-based representation, ensuring that generated specifications follow the expected structure and terminology. For behavioral parameterization, we build a module corpus $\mathcal{K}_{doc}$ covering 35 frequently used behavioral modules in Niagara.
The material library $\mathcal{M}$ consists of textual descriptions for 21 assets collected from the Fab marketplace \cite{Fab}, including static geometric primitives, sequential flipbook animations, and procedural materials with time-varying behaviors. To implement the DRAG mechanism, we utilize \textit{all-MiniLM-L6-v2} \cite{reimers2019sentence}  to facilitate semantic matching. We use a score threshold of $\tau = 0.9$ and a maximum of $T = 4$ iterations, which typically strikes a balance between visual quality and inference time.
\subsection{Quantitative Results}
\begin{figure*}[t]
  \centering

  \includegraphics[width=\linewidth]{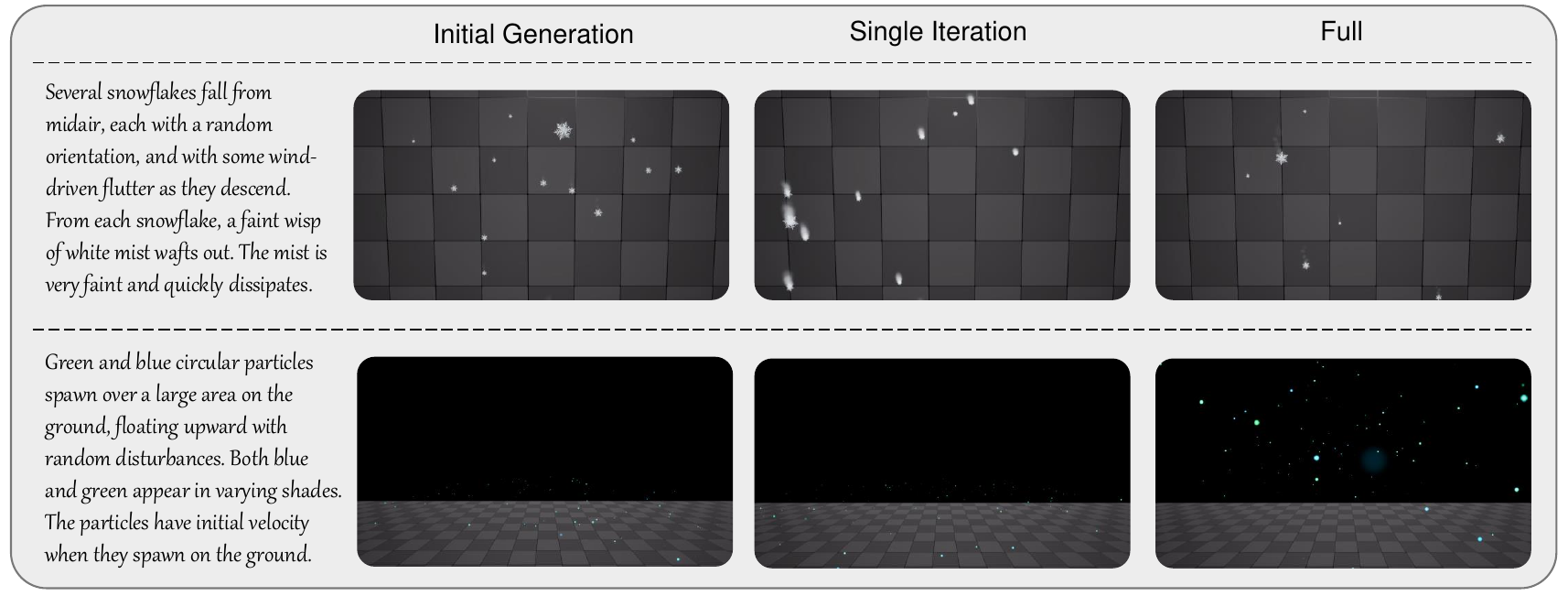}
  \caption{\textbf{Visual progression of the particle effect synthesis process across sequential stages.} For the snow effect (top), early stages produce mist that is either nearly invisible or excessively dense, and they struggle to correctly balance the wind-driven flutter. In the floating particles scenario (bottom), only full framework manifests obvious upward floating motion.}
  \label{fig:process}
\end{figure*}

We evaluate the generative capabilities of our framework by synthesizing a diverse range of particle effects from 75 distinct natural language descriptions $\mathcal{D}$. This evaluation dataset encompasses various categories including combat skill effects, natural phenomena, and fireworks, all constructed within the expressive capacity of our material library $\mathcal{M}$. Given the lack of existing generative baselines, we compare our full framework against the Initial generation (serving as the baseline) and single-iteration refinement.

The evaluation focuses on two dimensions: semantic alignment and visual aesthetics, with results reported in Table~\ref{tab:quantitative_evaluation_results}. Semantic alignment is measured using the CLIP4Clip model~\cite{luo2022clip4clip}, which provides objective video--text similarity scores. For each rendered result, we uniformly sample frames and encode them with a ViT-B/32 CLIP4Clip encoder. The resulting frame embeddings are mean-pooled to obtain a single video-level representation, and the CLIP4Clip score is computed as the cosine similarity between this representation and the text embedding of the user description $\mathcal{D}$.

To supplement the vector-based similarity, the evaluation adopts a Question Answering-based Vision-Language Model (VLM) evaluation protocol inspired by T2VScore~\cite{wu2024towards}. The protocol separates question generation from visual evaluation across the two aforementioned dimensions. For semantic alignment, a question-setting agent decomposes $\mathcal{D}$ into multiple-choice questions with reference answers. For visual aesthetics, we design 1-5 rating questions drawing on the visual-output quality dimensions in the KinemaFX questionnaire~\cite{zhang2025kinemafx}, such as kinematic harmony and effect artwork quality.
A visual-evaluation agent observes sampled frames, answers the semantic alignment questions using only visible evidence, and scores the visual aesthetics rating questions. Our full framework received an average semantic matching score of 4.200 and an aesthetic score of 4.043, demonstrating a substantial enhancement over the initial generation baseline. We use GPT-5.4 Mini~\cite{gpt} as the question-setting agent, while Gemini 3.1 Pro~\cite{gemini} serves as both the visual-evaluation agent and an independent evaluator, ensuring impartiality and avoiding bias relative to our \textit{Critic agent}.

\begin{table}[t]
  \centering
  \small
  \setlength{\tabcolsep}{2.5pt}
  \caption{\textbf{Quantitative evaluation across different frameworks.} Metrics are categorized into semantic alignment and visual aesthetics. Our full pipeline consistently achieves the highest performance across both automated metrics (CLIP4Clip, VLM) and human evaluation (User).}
  \label{tab:quantitative_evaluation_results}
    \vspace{-5pt}
  \begin{tabular}{@{}lccccc@{}}
    \toprule
    \multirow{2}{*}{\textbf{Configuration}} & \multicolumn{3}{c}{\textbf{Semantic Alignment}} & \multicolumn{2}{c}{\textbf{Visual Aesthetics}} \\
    \cmidrule(lr){2-4} \cmidrule(lr){5-6}
    & \textbf{CLIP4Clip} $\uparrow$ & \textbf{VLM} $\uparrow$ & \textbf{User} $\uparrow$ & \textbf{VLM} $\uparrow$ & \textbf{User} $\uparrow$ \\
    \midrule
    Initial Gen.         & 0.267 & 3.259 & 3.255 & 3.570 & 3.190 \\
    Single-iter.          & 0.273 & 3.586 & 3.680 & 3.492 & 3.620 \\
    \textbf{Full (Ours)} & \textbf{0.281} & \textbf{4.200} & \textbf{4.420} & \textbf{4.043} & \textbf{4.400}\\
    \bottomrule
  \end{tabular}
\end{table}
The validity of these automated metrics is further reinforced by a user study involving 20 professional technical artists. In a blind evaluation of 10 randomly sampled effects, the experts consistently assigned the highest ratings to the full framework, with mean scores of $4.420$ for semantic alignment and $4.400$ for visual aesthetics. The quantitative results show a slight divergence between VLM scores and human ratings regarding visual aesthetics during the early refinement stages. This discrepancy stems from the fact that the \textit{Refiner} primarily focuses on correcting semantic misalignments.
Human evaluators often conflate the resulting behavioral accuracy with visual appeal, leading them to reward the fulfillment of the description with higher aesthetic ratings. Conversely, the VLM evaluates aesthetics based on a fixed set of visual criteria that remain decoupled from the semantic prompt.

\subsection{Qualitative Results}

\begin{figure*}[t]
  \includegraphics[width=\linewidth]{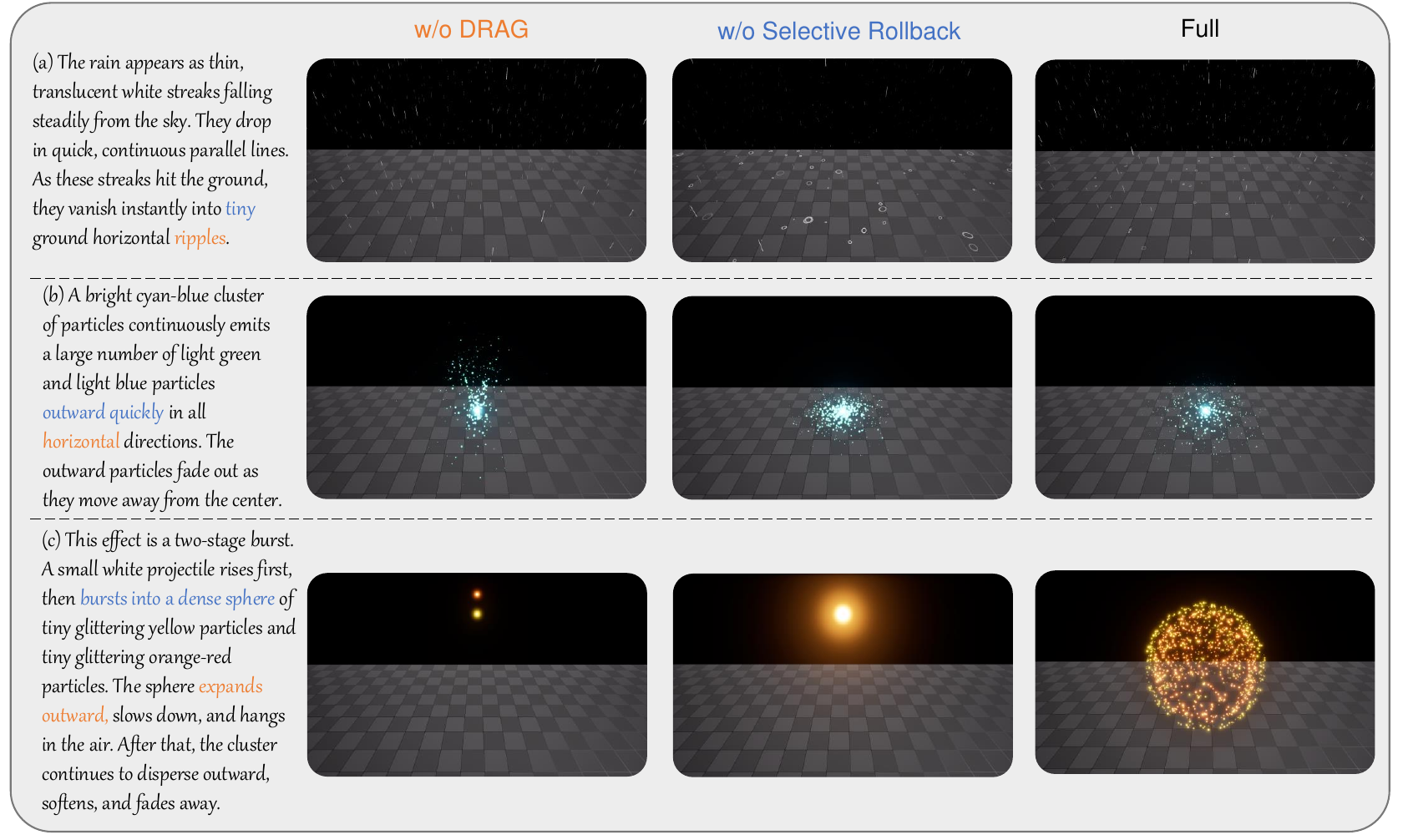}
  \caption{\textbf{Visual ablation study of the core iterative refinement components.}
  We compare our full framework against two variants: one without the DRAG mechanism and one without selective rollback. Text highlights correspond to failures in the respective columns: orange denotes causal misattribution without DRAG, and blue denotes effects degraded without selective rollback.
  (a) Rain: Without DRAG, the system misses the dependency where the \textit{Collision} module resets velocity and suppresses the \textit{ripple} trigger; without rollback, it retains the oversized ripple scale adjusted while the effect was invisible.
  (b) Horizontal spray: The system fails to correct the \textit{Shape Location} axis without DRAG, while the lack of rollback leaves the velocity unnecessarily low.
  (c) Two-stage burst: Without DRAG, the agent fails to diagnose that the position set by \textit{Shape Location} is overwritten by event data, and without rollback, the result retains the erroneously lowered velocity.}
  \label{fig:ablation}
\end{figure*}

\subsubsection{Generated Results}
\begin{figure*}[t]
  \centering
  \includegraphics[width=\linewidth]{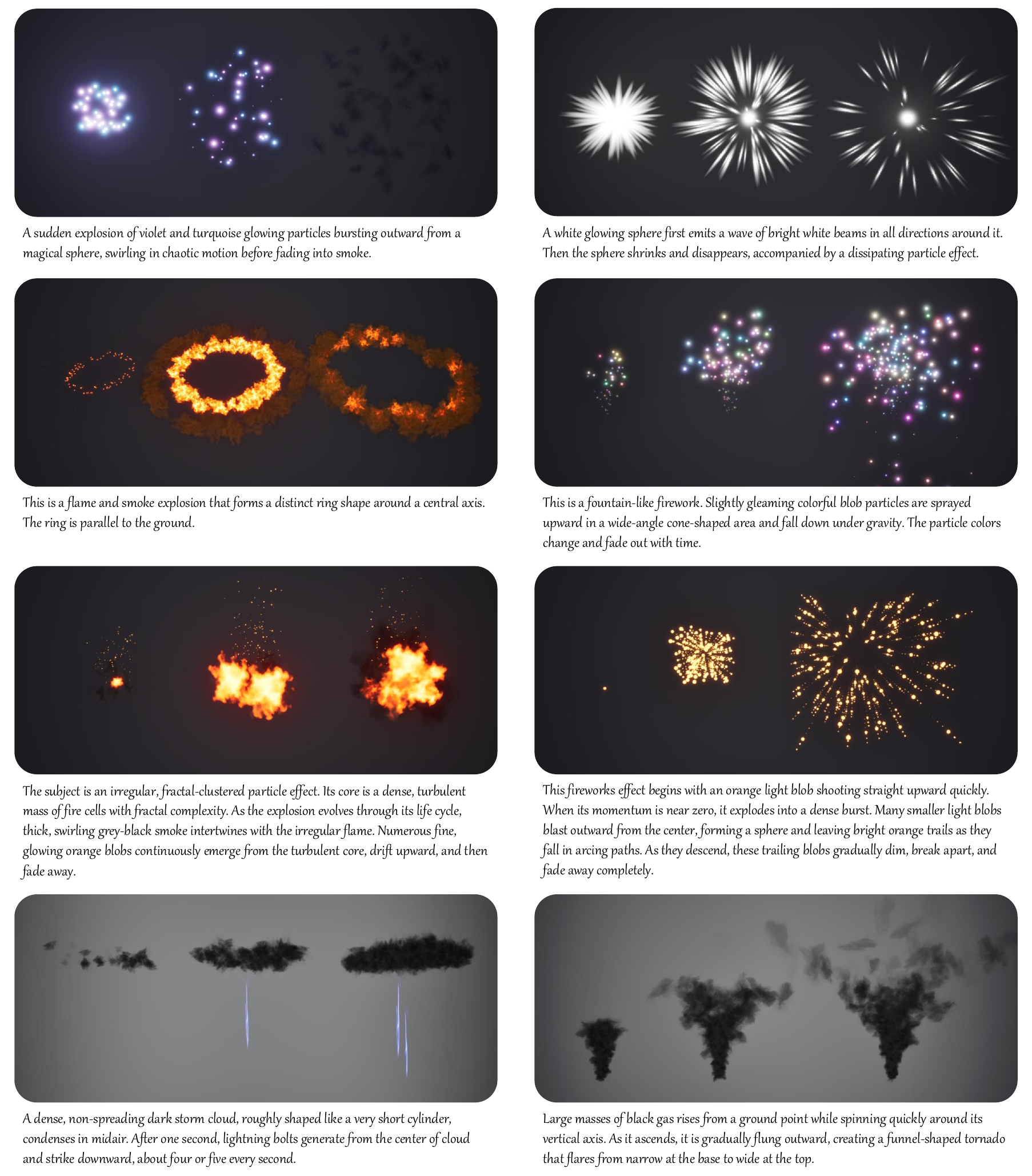}
  \caption{\textbf{Representative examples of particle effects synthesized using our framework.} Each effect is decomposed into several phases, displayed left to right to show the simulation's progression over time. These results showcase the system's ability to generate diverse visual phenomena encompassing elemental spells, natural phenomena, and fireworks.}
  \label{fig:gallery}
\end{figure*}

Fig.~\ref{fig:showcase} presents three representative particle effects synthesized by our full framework, spanning categories such as fireworks and abstract magical spells. The synthesized effects utilize the expressive capacity of the Niagara~\cite{niagara} framework. These results demonstrate the system's ability to map complex linguistic descriptions to expressive simulation logic. More results are shown in Fig.~\ref{fig:gallery}. Since our system produces fully editable particle systems, users can perform secondary manual adjustments to the synthesized assets, enabling the realization of more sophisticated visual phenomena that align precisely with specific creative or production requirements. This adaptability ensures that our generative framework serves as a powerful starting point in the professional particle effects pipeline, allowing for further artistic refinement beyond the automated synthesis.

\subsubsection{Ablation Study.}

The visual progression across the synthesis stages is illustrated in Fig.~\ref{fig:process}. The initial generation stage establishes the core structural layout and basic behavioral logic, but it often lacks fine-grained alignment with nuanced artistic cues. In the snow simulation, for instance, the initial mist is barely visible, and the specified wind-driven flutter is absent. Furthermore, due to the high-dimensional complexity of the simulation graph, the initial synthesis occasionally overlooks critical inter-module dependencies, leading to unexpected behaviors such as broken motion trajectories or incomplete rendering states.

A single refinement iteration can rectify obvious discrepancies, such as base colors or spawn rates, but often fails to achieve the required precision across all requested behaviors in one pass. In the snow example, the mist becomes excessively dense after one refinement iteration. Similarly, a single iteration may fail to resolve complex motion coupling, as shown in the second scenario, where particles remain clustered on the ground. In contrast, our full framework achieves more accurate behavioral alignment and more sophisticated motion by iteratively resolving these dependencies through closed-loop refinement.

The contributions of DRAG and selective rollback are further validated by the visual comparison in Fig.~\ref{fig:ablation}. Without DRAG, the \textit{Refiner agent} is prone to causal misattribution. As shown in the rain simulation example in Fig.~\ref{fig:ablation}(a), the \textit{Refiner agent} fails to generate rain ripples without DRAG because the \textit{Collision} module resets particle velocity to zero upon collision. This prevents the velocity threshold required by the \textit{Generate Collision Event} module from being met and thereby suppresses secondary ripple effects.

The selective rollback strategy serves as a safeguard against parameter drift during iterative refinement. In the rain simulation, the ripples are initially invisible due to the logic error described above. In early refinement attempts, the \textit{Refiner agent} may therefore try to compensate by incrementally increasing the \textit{spawn rate} or \textit{scale} over multiple iterations. Although each individual modification may be subtle, these ungrounded adjustments can accumulate over time. Without rollback, the effects of previous parameter changes remain in the system even after the agent identifies the correct logic fix. These residual adjustments often require additional iterations to correct, and may occasionally remain unresolved. Furthermore, such uncorrected changes increase the complexity of the simulation state, making it more difficult to identify the underlying causes of failure.

\section{Discussion}
\paragraph{Limitations.}
The practical application of our framework is mainly constrained by three factors. Primarily, the system is susceptible to LLM context window limits as the refinement process necessitates a comprehensive representation of all parameters for the \textit{Refiner} to evaluate complex dependencies between modules and emitters. This can lead to reasoning degradation when complex effects involve a large number of coordinated emitters. Secondly, the creative range is bounded by a predefined material library. While the framework is theoretically applicable to procedural material synthesis, it requires the integration of texture generation workflows to achieve fully open-ended expressivity. Lastly, natural language descriptions are inherently limited in achieving fine-grained control. Vague semantic prompts are often insufficient to specify exact physical properties or precise spatial layouts, making it difficult to meet the rigorous precision required for professional 3D production.
\paragraph{Future Works.}
Future research could investigate the end-to-end synthesis of particle systems, where simulation logic and materials are generated simultaneously. This would remove the dependency on pre-existing libraries and significantly broaden the expressive power of the synthesis framework. Additionally, the multi-agent architecture could be leveraged for the automated validation and optimization of existing particle assets. Specialized agents could be developed to audit technical integrity and refine the computational performance of complex simulations, ensuring their suitability for diverse runtime environments. Another promising direction involves the integration of multimodal inputs, including hand-drawn sketches or reference video sequences, to provide more granular control over the generative process. By incorporating visual cues, the framework could better resolve spatial and temporal constraints that natural language descriptions often struggle to specify.

\section{Conclusions}
In this paper, we introduced ParticleGen, a multi-agent framework for structured synthesis of real-time particle effects from natural-language descriptions. By decomposing the generation process into structural planning, behavioral parameterization, and visual-feedback refinement, our approach bridges high-level artistic intent and low-level particle-system specifications. The iterative refinement loop, supported by a specialized diagnostic RAG mechanism, improves the alignment between synthesized effects and user descriptions by identifying visual artifacts and tracing them to procedural causes. Our evaluation within UE5's Niagara system demonstrates that ParticleGen can generate diverse, high-quality, and editable particle effects in an engine-native workflow. By moving beyond retrieval-based effect composition toward direct synthesis of simulation logic and parameters, this work suggests a practical path toward more accessible and efficient particle-effect authoring.

\bibliography{references}

\end{document}